\documentstyle[12pt]{article}
\begin{document}

\title{\Large\bf Hamiltonian embedding of the massive Yang-Mills theory and the
generalized St\"uckelberg formalism}

\author{\\
R. Banerjee\thanks{\noindent e-mail: rabin@if.ufrj.br --
Permanent address: S.N. Bose National Centre for Basic Sciences, Block JD,
Sector III, Salt Lake, Calcutta- 700091 , India.}~ and J.
Barcelos-Neto\thanks{\noindent 
e-mail: barcelos@if.ufrj.br}\\ 
Instituto de F\'{\i}sica\\
Universidade Federal do Rio de Janeiro\\ 
RJ 21945-970 - Caixa Postal 68528 - Brasil\\}
\date{}

\maketitle
\abstract
Using the general notions of Batalin, Fradkin, Fradkina and Tyutin to
convert second class systems into first class ones, we present a
gauge invariant formulation of the massive Yang-Mills theory by
embedding it in an extended phase space. The infinite set of
correction terms necessary for obtaining the involutive constraints
and Hamiltonian is explicitly computed and expressed in a closed form.
It is also shown that the extra fields introduced in the correction
terms are exactly identified with the auxiliary scalars used in the
generalized St\"uckelberg formalism for converting a gauge
noninvariant Lagrangian into a gauge invariant form.

\vfill\noindent

\vspace{1cm}
\newpage
\section{Introduction}

\bigskip
The foundations for a Hamiltonian formulation of constrained systems
were originally laid down by Dirac \cite{D} whose work still remains
fundamental to our understanding of the subject. For second class
systems, however, this approach poses problems which are related to
ambiguities that may arise in the transition of (classical) Dirac
brackets to (quantum) commutators. During recent years a viable and
powerful alternative for discussing such systems has been developed
in a series of papers by Batalin, Fradkin, Fradkina and Tyutin (BFFT)
\cite{B,BT}. The basic idea is to convert, using an iterative
prescription, the second class system into a first class one by
enlarging the original phase space. Once this is achieved it is
possible to exploit the existing machinery available for quantising
first class theories \cite{HT}.  The use of Dirac brackets is
completely avoided.

\medskip
The BFFT method, or modified versions of it \cite{BR,BB,A,K}, have
been used by several authors \cite{X} to discuss the formulation of
specific examples of second class theories. These have in turn
provided an insight into the general formalism which cannot be
otherwise gained. But such analyses has been mostly confined to
Abelian models where the constraints are linear. For nonlinear
constraints which occur naturally in non-Abelian theories, the
situation is rather complicated and explicit computations \cite{BBG}
are quite sparse and, at best, limited to a simple group like
$SU(2)$. The reason for this complication is that, contrary to the
example of linear constraints, the iterative process in the BFFT
approach may not terminate. Nevertheless it was shown by one of us
\cite{BBG1} that the nonlinear constraints occurring in the nonlinear
sigma models could be tackled by a clever choice of the generating
matrices which forces the iterative process to stop at the first
step. Unfortunately, such a simplification cannot be done in general
\cite{BN}.  Then the iterative scheme has to be continued
indefinitely in which case it is not clear whether consistent
solutions for all the steps can be obtained from which meaningful
closed form expressions for the involutive constraints and
Hamiltonian may be determined.

\medskip
In this paper we systematically develop an algorithm within the BFFT
approach by which all the iterative corrections necessary to
transform the original second class nonlinear constraints into
strongly involutive constraints can be explicitly computed and
exhibited in a closed form. We shall discuss this in Section 3 within
the context of the massive Yang-Mills theory which is a classic
example of a second class system with nonlinear constraints.  The
iterative corrections are explicitly computed and form an infinite set
which is expressed in a closed (exponential-like) form. We then use a
modified prescription, which is a departure from the conventional
BFFT ideas, of computing the involutive Hamiltonian.  Once again the
iterative terms can be combined to yield an exponential-like series.
The modified prescription mentioned above provides great technical
simplifications. This has been outlined in Section 2 which also
contains a short review of the BFFT approach.  In our entire analysis
there is no restriction on either the dimensionality of space time or
the specific non-Abelian gauge group employed in the model. Since the
gauge invariance of the massive Yang-Mills theory is broken by the
mass term, it will be clear that our analysis is applicable to all
such generic second class systems whose lack of gauge invariance is
attributed to identical reasons.  In Section 4 we first reconsider
the generalized St\"uckelberg formalism \cite{KG} of converting the
massive Yang-Mills Lagrangian into a gauge invariant form by the
introduction of auxiliary scalars.  Next, the canonical formalism of
the gauge invariant Lagrangian is developed in details.  Finally, by
a series of remarkable algebraic manipulations it is shown that these
scalars are exactly identified with the additional fields that occur
in the correction terms found in the BFFT approach.  This establishes
a one-to-one correspondence between the Lagrangian and Hamiltonian
embedding prescriptions of converting gauge noninvariant into gauge
invariant systems. Some appendices are included for elaborating
calculations relevant for Section 4. Our conclusions are given in
Section~5.

\vspace{1cm}
\section{Brief review of the BFFT formalism}
\renewcommand{\theequation}{2.\arabic{equation}}
\setcounter{equation}{0}

\bigskip
In this section we present a short review of the Hamiltonian
formalism, which will be relevant for the subsequent discussion, for
converting second class systems into true gauge (i.e. first class)
systems by enlarging the phase space. Towards the end we shall also
discuss a deviation from the conventional approach that provides
considerable technical simplification.

\medskip
Let us take a system described by a Hamiltonian $H_0$ in a
phase-space with variables $(q^i, p_i)$ where $i$ runs from 1 to N.
For simplicity these variables are assumed to be bosonic (extension
to include fermionic degrees of
freedom and to the continuous case
can be done in a straightforward
way). It is also supposed that
there  exists second-class
constraints only since this is the case
that will be investigated.  Denoting them by $T_a$, with
$a=1,\dots,M<2N$, we have

\begin{equation}
\bigl\{T_a,\,T_b\bigr\}=\Delta_{ab}\,,
\label{2.1}
\end{equation}

\bigskip\noindent
where $\det(\Delta_{ab})\not=0$. 

\medskip 
The first objective is to transform these second-class constraints
into first-class ones.  Towards this goal auxiliary variables
$\eta^a$ are introduced, one for each second-class constraint (the
connection between the number of constraints and the new variables in
a one-to-one correlation is to keep the same number of the physical
degrees of freedom in the resulting extended theory), which satisfy a
symplectic algebra,

\begin{equation}
\label{2.2}
\bigl\{\eta^a,\,\eta^b\bigr\}=\omega^{ab}\,,
\end{equation}

\bigskip\noindent
where $\omega^{ab}$ is a constant quantity with
$\det\,(\omega^{ab})\not=0$.  The first class constraints are now
defined by,

\begin{equation}
\label{2.3}
\tilde T_a=\tilde T_a(q,p;\eta)\,,
\end{equation}

\bigskip\noindent 
and satisfy the boundary condition

\begin{equation}
\label{2.4}
\tilde T_a(q,p;0)=T_a(q,p)\,.
\end{equation}

\bigskip\noindent
A characteristic of these new constraints is that they are assumed to
be strongly involutive,i.e.

\begin{equation}
\label{2.5}
\bigl\{\tilde T_a,\,\tilde T_b\bigr\}=0\,.
\end{equation}

\bigskip
The solution of~(\ref{2.5}) can be achieved by considering an
expansion of $\tilde T_a$, as

\begin{equation}
\label{2.6}
\tilde T_a=\sum_{n=0}^\infty T_a^{(n)}\,,
\end{equation}

\bigskip\noindent
where $T_a^{(n)}$ is a term of order $n$ in $\eta$.  Compatibility
with the boundary condition~(\ref{2.4}) requires that

\begin{equation}
\label{2.7}
T_a^{(0)}=T_a\,.
\end{equation}

\bigskip\noindent
The replacement of~(\ref{2.6}) into~(\ref{2.5}) leads to a set of
recursive relations, one for each coefficient of $\eta^n$. We
explicitly list the equations for $n =0, 1, 2$,

\begin{eqnarray}
\bigl\{T_a^{(0)},T_b^{(0)}\bigr\}_{(q,p)}
&\!\!\!\!+\!\!\!\!&\bigl\{T_a^{(1)},T_b^{(1)}\bigr\}_{(\eta)}=0\,,
\label{2.8a}\\
\bigl\{T_a^{(0)},T_b^{(1)}\bigr\}_{(q,p)}
&\!\!\!\!+\!\!\!\!&\bigl\{T_a^{(1)},T_b^{(0)}\bigr\}_{(q,p)}
+\bigl\{T_a^{(1)},T_b^{(2)}\bigr\}_{(\eta)}
+\bigl\{T_a^{(2)},T_b^{(1)}\bigr\}_{(\eta)}=0\,,
\label{2.8b}\\
\bigl\{T_a^{(0)},T_b^{(2)}\bigr\}_{(q,p)}
&\!\!\!\!+\!\!\!\!&\bigl\{T_a^{(1)},T_b^{(1)}\bigr\}_{(q,p)}
+\bigl\{T_a^{(2)},T_b^{(0)}\bigr\}_{(q,p)}
+\bigl\{T_a^{(1)},T_b^{(3)}\bigr\}_{(\eta)}
\nonumber\\
&\!\!\!\!+\!\!\!\!&\bigl\{T_a^{(2)},T_b^{(2)}\bigr\}_{(\eta)}
+\bigl\{T_a^{(3)},T_b^{(1)}\bigr\}_{(\eta)}=0\,,
\label{2.8c}\\
&\vdots&\nonumber
\end{eqnarray}

\bigskip\noindent
The notations $\{,\}_{(q,p)}$ and $\{,\}_{(\eta)}$ used above
represent the parts of the Poisson bracket $\{,\}$ relative to the
variables $(q,p)$ and
$(\eta)$.
\footnote{Sometimes in explicit calculations if no suffix appears in
the definition of the Poisson bracket, it will imply an evaluation
relative to the initial phase space variables ($q,p$).}

\medskip
The above equations are used iteratively to obtain the corrections
$T^{(n)}$ ($n\geq1$).  Equation~(\ref{2.8a}) shall give $T^{(1)}$.
With this result and~(\ref{2.8b}), one calculates $T^{(2)}$, and so
on. Since $T^{(1)}$ is linear in $\eta$ we may write

\begin{equation}
\label{2.9}
T_a^{(1)}=X_{ab}(q,p)\,\eta^b\,.
\end{equation}

\bigskip\noindent
Introducing this expression into~(\ref{2.8a}) and using the
boundary
condition~(\ref{2.4}), as well as~(\ref{2.1})
and~(\ref{2.2}), we get

\begin{equation}
\label{2.10}
\Delta_{ab}+X_{ac}\,\omega^{cd}\,X_{bd}=0\,.
\end{equation}

\bigskip\noindent
We notice that this equation  contains two  unknowns $X_{ab}$ and
$\omega^{ab}$. Usually, first of all $\omega^{ab}$ is chosen in such
a way that the new variables are unconstrained. It is opportune to
mention that it is not always possible to make such a choice
\cite{AB}. In consequence, the consistency of the method requires an
introduction of other new variables in order to transform these
constraints also into first-class. This may lead to an endless
process. However, it is important to emphasize that $\omega^{ab}$ can
be fixed anyway.

\medskip
After fixing $\omega^{ab}$, we pass to consider the coefficients
$X_{ab}$. They cannot be obtained  unambiguously since, even after
fixing $\omega^{ab}$, expression (\ref{2.10}) leads to less equations
than variables. The choice of $X$'s has therefore to be done in a
convenient way \cite{BBG1}.

\medskip
The knowledge of $X_{ab}$ permits us to obtain $T_a^{(1)}$. If
$X_{ab}$ does not depend on $(q,p)$, it is easily seen that
$T_a+T_a^{(1)}$ is already strongly involutive and we succeed in
obtaining $\tilde T_a$.  This is what happens for systems with linear
constraints. For nonlinear constraints, on the other hand, $X_{ab}$
becomes variable dependent which necessitates the analysis to be
pursued beyond the first iterative step.  All the subsequent
corrections must be explicitly computed, the knowledge of $T_a^{(n)}
(n=0, 1, 2,...n) $ leading to the evaluation of $T_a^{(n+1)}$ from
the recursive relations.  Once again the importance of choosing the
proper  solution for $X_{ab}$ becomes apparent otherwise the series
of corrections cannot be put in a closed form and the expression for
the involutive constraints becomes unintelligible and uninteresting.

\medskip
Another point in the Hamiltonian formalism is that any dynamic
function $A(q,p)$ (for instance, the Hamiltonian) has also to be
properly modified in order to be strongly involutive with the
first-class constraints $\tilde T_a$. Denoting the modified quantity
by $\tilde A(q,p;\eta)$, we then have

\begin{equation}
\label{2.11}
\bigl\{\tilde T_a,\,\tilde A\bigr\}=0\,.
\end{equation}

\bigskip\noindent
In addition, $\tilde A$ has also to satisfy  the boundary
condition

\begin{equation}
\label{2.12}
\tilde A(q,p;0)=A(q,p)\,.
\end{equation}

\bigskip
To obtain $\tilde A$ an expansion analogous to (\ref{2.6}) is
considered,

\begin{equation}
\label{2.13}
\tilde A=\sum_{n=0}^\infty A^{(n)}\,,
\end{equation}

\bigskip\noindent 
where $A^{(n)}$ is also a term of order $n$ in $\eta$'s.
Consequently, compatibility with~(\ref{2.12}) requires that

\begin{equation}
\label{2.14}
A^{(0)}=A\,.
\end{equation}

\bigskip\noindent
The combination of~(\ref{2.6}),~(\ref{2.11}) and~(\ref{2.13}) gives

\begin{eqnarray}
&&\bigl\{T_a^{(0)},A^{(0)}\bigr\}_{(q,p)}
+\bigl\{T_a^{(1)},A^{(1)}\bigr\}_{(\eta)}=0\,,
\label{2.15a}\\
&&\bigl\{T_a^{(0)},A^{(1)}\bigr\}_{(q,p)}
+\bigl\{T_a^{(1)},A^{(0)}\bigr\}_{(q,p)}
+\bigl\{T_a^{(1)},A^{(2)}\bigr\}_{(\eta)}
\nonumber\\
&&\phantom{\bigl\{T_a^{(0)},A^{(0)}\bigr\}_{(q,p)}}
+\bigl\{T_a^{(2)},A^{(1)}\bigr\}_{(\eta)}=0\,,
\label{2.15b}\\
&&\bigl\{T_a^{(0)},A^{(2)}\bigr\}_{(q,p)}
+\bigl\{T_a^{(1)},A^{(1)}\bigr\}_{(q,p)}
+\bigl\{T_a^{(2)},A^{(0)}\bigr\}_{(q,p)}
\nonumber\\
&&\phantom{\bigl\{T_a^{(0)},A^{(0)}\bigr\}_{(q,p)}}
+\bigl\{T_a^{(1)},A^{(3)}\bigr\}_{(\eta)}
+\bigl\{T_a^{(2)},A^{(2)}\bigr\}_{(\eta)}
\nonumber\\
&&\phantom{\bigl\{T_a^{(0)},A^{(0)}\bigr\}_{(q,p)}}
+\bigl\{T_a^{(3)},A^{(1)}\bigr\}_{(\eta)}=0\,,
\label{2.15c}\\
&&\phantom{\bigl\{T_a^{(0)},A^{(0)}\bigr\}_{(q,p)}+}
\vdots
\nonumber
\end{eqnarray}

\bigskip\noindent
which correspond to the coefficients of the powers $\eta^0$,
$\eta^1$, $\eta^2$, etc., respectively. The expression~(\ref{2.15a})
above gives us $A^{(1)}$

\begin{equation}
\label{2.16}
A^{(1)}=-\,\eta^a\,\omega_{ab}\,X^{bc}\,
\bigl\{T_c,\,A\bigr\}\,,
\end{equation}

\bigskip\noindent
where $\omega_{ab}$ and $X^{ab}$ are the inverses of $\omega^{ab}$
and $X_{ab}$.

\medskip
It was earlier seen that $T_a+T^{(1)}_a$ was strongly involutive if
the coefficients $X_{ab}$ do not depend on $(q,p)$. However, the same
argument does not necessarily apply in this case. Usually we have to
calculate other corrections to obtain the final $\tilde A$. Let us
discuss how this can be systematically done. We consider the general
case first.  The correction $A^{(2)}$ comes from
equation~(\ref{2.15b}), that we conveniently rewrite as

\begin{equation}
\label{2.17}
\bigl\{T_a^{(1)},\,A^{(2)}\bigr\}_{(\eta)}
=-\,G_a^{(1)}\,,
\end{equation}

\bigskip\noindent
where

\begin{equation}
\label{2.18}
G_a^{(1)}=\bigl\{T_a,\,A^{(1)}\bigr\}_{(q,p)}
+\bigl\{T_a^{(1)},\,A\bigr\}_{(q,p)}
+\bigl\{T_a^{(2)},\,A^{(1)}\bigr\}_{(\eta)}\,.
\end{equation}
\bigskip\noindent

Thus

\begin{equation}
\label{2.19}
A^{(2)}=-{1\over2}\,\eta^a\,\omega_{ab}\,X^{bc}\,G_c^{(1)}\,.
\end{equation}

\bigskip\noindent
In the same way, other terms can be obtained. The final general
expression reads

\begin{equation}
\label{2.20}
A^{(n+1)}=-{1\over n+1}\,\eta^a\,\omega_{ab}\,X^{bc}\,G_c^{(n)}\,,
\end{equation}

\bigskip\noindent
where

\begin{equation}
\label{2.21}
G_a^{(n)}=\sum_{m=0}^n\bigl\{T_a^{(n-m)},\,A^{(m)}\bigr\}_{(q,p)}
+\sum_{m=0}^{n-2}\bigl\{T_a^{(n-m)},\,A^{(m+2)}\bigr\}_{(\eta)}
+\bigl\{T_a^{(n+1)},\,A^{(1)}\bigr\}_{(\eta)}\,.
\end{equation}

\bigskip
Although it is possible to convert any dynamical variable in the
original phase space into its involutive form by the above method,
there may be technical problems in carrying out this construction
particularly if one considers non-Abelian theories. In such cases the
relevant variable is already quite complicated and this process may
lead to an arcane structure which would not be illuminating. We
suggest the following simplification which has been employed, though
in a modified form, earlier for abelian models \cite{BB,K}. The basic
idea is to obtain the involutive forms of the initial fields $q$ and
$p$. This can be derived from the previous analysis. Denoting these
by $\tilde q$ and $\tilde p$ so that,

\begin{equation}
\bigl\{\tilde T,\,\tilde q\bigr\}
=\bigl\{\tilde T,\,\tilde p\bigr\}= 0\,.
\label{2.26}
\end{equation}

\bigskip\noindent
Now any function of $\tilde q$ and $\tilde p$ will also be strongly
involutive since,

\begin{eqnarray}
\bigl\{\tilde T,\,\tilde F(\tilde q, \tilde p)\bigr\}
&=&\bigl\{\tilde T,\,\tilde q\bigr\}\,
\frac{\partial\tilde F}{\partial\tilde q}
+\bigl\{\tilde T,\,\tilde p\bigr\}\,
\frac{\partial\tilde F}{\partial\tilde p}\,,
\nonumber\\
&=&0\,.
\label{2.27}
\end{eqnarray}

\bigskip\noindent
Thus if we take any dynamical variable in the original phase space,
its involutive form can be obtained by the replacement,

\begin{equation}
F(q, p) \rightarrow F(\tilde q, \tilde p)
=\tilde F(\tilde q,\tilde p)\,.
\label{2.28}
\end{equation}

\bigskip\noindent
It is obvious that the initial boundary condition in the BFFT
process, namely, the reduction of the involutive function to the
original function when the new fields are set to zero, remains
preserved.

\vspace{1cm}
\section{Involutive constraints and Hamiltonian in the massive
Yang-Mills theory}
\renewcommand{\theequation}{3.\arabic{equation}}
\setcounter{equation}{0}

\bigskip
The massive Yang-Mills theory is defined by the Lagrangian density,

\begin{equation}
{\cal L}=-\,\frac{1}{4}\,F_{\mu\nu}^aF^{a\mu\nu}
+\frac{1}{2}\,m^2\,A_\mu^aA^{a\mu}\,,
\label{3.1}
\end{equation}

\bigskip\noindent
where the following conventions and notations will be used,

\begin{eqnarray}
F_{\mu\nu}^a&=&\partial_\mu A_\nu^a
-\partial_\nu A_\mu^a+gf^{abc}\,A_\mu^bA_\nu^c\,,
\nonumber\\
\bigl[T^a,\,T^b\bigr]&=&i\,f^{abc}\,T^c\,,
\nonumber\\
{\rm tr}\,\bigl(T^aT^b\bigr)&=&\frac{1}{2}\,\delta^{ab}\,,
\nonumber\\
\bigl(T^a\bigr)^{bc}&=&i\,f^{abc}\,.
\label{3.2}
\end{eqnarray}
\bigskip\noindent

The canonical momentum conjugated to $A_\mu^a$ reads,

\begin{equation}
\pi_\mu^a=\frac{\partial{\cal L}}{\partial\dot A^{a\mu}}
=-\,F_{0\mu}^a\,.
\label{3.3}
\end{equation}

\bigskip\noindent
Hence, $\pi_0^a$ is a primary constraint,

\begin{equation}
T_1^a=\pi_0^a\approx0\,.
\label{3.4}
\end{equation}

\bigskip\noindent
In order to look for secondary constraints, we construct the
canonical Hamiltonian density,

\begin{eqnarray}
{\cal H}_c&=&\pi^{a\mu}\dot A_\mu^a-{\cal L}\,,
\nonumber\\
&=&\pi^{a0}\dot A_0^a-\frac{1}{2}\,\pi^{ai}\pi^a_i
+\pi^{ai}\partial_i A^a_0-gf^{abc}\,\pi^{ai}A_0^bA_i^c
\nonumber\\
&&+\frac{1}{4}\,F_{ij}^aF^{aij}-\frac{1}{2}\,m^2\,A^a_0A^{a0}
-\frac{1}{2}\,m^2\,A^a_iA^{ai}\,.
\label{3.5}
\end{eqnarray}

\bigskip\noindent
The total Hamiltonian \cite{D} is given by,

\begin{equation}
H_T=\int d^3x\,\Bigl({\cal H}_c+\lambda^a\,\pi_0^a\Bigr)\,.
\label{3.6}
\end{equation}

\bigskip\noindent
In Eq. (\ref{3.6}) note that the term $\pi^{a0}\dot A_0^a$ that
appears in (\ref{3.5}) has been absorbed in the $\lambda^a\,\pi_0^a$
term by a redefinition of the Lagrange multiplier $\lambda^a$.
Consequently, ${\cal H}_c$ occurring in (\ref{3.6}) differs from
(\ref{3.5}) by this piece. The consistency condition for the primary
constraint,

\begin{eqnarray}
\bigl\{\pi_0^a(x),\,H_T\bigr\}&=&\partial_i\pi^{ai}
+gf^{abc}\,A_i^b\,\pi^{ci}+m^2\,A_0^a\,,
\nonumber\\
&=&(D_i\pi^i)^a+m^2\,A_0^a\approx0\,.
\label{3.7}
\end{eqnarray}

\bigskip\noindent
yields a secondary constraint,

\begin{equation}
T_2^a=(D_i\pi^i)^a+m^2\,A_0^a\approx0\,.
\label{3.8}
\end{equation}

\bigskip\noindent
There are no more constraints since the Poisson algebra
\footnote{All algebra will be implemented at equal times.
Furthermore, the three dimensional delta function
$\delta^3(\vec x-\vec y)$ will be written simply as $\delta(x-y)$.}
of the constraints $T_1^a$ and $T_2^a$ is noninvolutive,

\begin{eqnarray}
\bigl\{T_1^a(x),\,T_1^b(y)\bigr\}&=&0\,,
\label{3.9}\\
\bigl\{T_1^a(x),\,T_2^b(y)\bigr\}
&=&-\,m^2\,\delta^{ab}\,\delta(x-y)\,,
\label{3.10}\\
\bigl\{T_2^a(x),\,T_2^b(y)\bigr\}
&=&gf^{abc}\,(D_i\pi^i)^c\,\delta(x-y)\,.
\label{3.11}
\end{eqnarray}

\bigskip\noindent
The quantities $\Delta^{ab}_{AB}(x,y)\,\,\,(A,B=1,2)$ introduced in
Eq.(\ref{2.1}) are therefore given by,

\begin{eqnarray}
&&\Delta^{ab}_{11}(x,y)=0\,,
\nonumber\\
&&\Delta^{ab}_{12}(x,y)=-\,m^2\,\delta^{ab}(x-y)
=-\,\Delta^{ab}_{21}(x,y)\,,
\nonumber\\
&&\Delta^{ab}_{22}(x,y)=g\,f^{abc}\,(D_i\pi^i)^c\,\delta(x-y)\,.
\label{3.12}
\end{eqnarray}

\bigskip
Let us now extend the phase space by introducing the set of new
variables $(\eta^{1a},\,\eta^{2a})$. We consider them as canonical,
i.e.,

\begin{eqnarray}
&&\bigl\{\eta^{1a}(x),\,\eta^{2b}(y)\bigr\}
=\delta^{ab}\,\delta(x-y)\,,
\nonumber\\
&&\bigl\{\eta^{1a}(x),\,\eta^{1b}(y)\bigr\}=0
=\bigl\{\eta^{2a}(x),\,\eta^{2b}(y)\bigr\}\,.
\label{3.13}
\end{eqnarray}

\bigskip\noindent
The symplectic matrix $\omega$ defined in Eq. (\ref{2.2}) therefore
has the structure,

\begin{equation}
\Bigl(\omega^{aA\,bB}(x,y)\Bigr)
=\left(\begin{array}{cc}
0&1\\
-\,1&0
\end{array}\right)\,\delta^{ab}\,\delta(x-y)\,.
\label{3.14}
\end{equation}

\bigskip\noindent
To calculate the first correction of the constraints we have to solve
Eq. (\ref{2.10}), which is now written as,

\begin{equation}
\Delta_{AB}^{ab}(x,\,y)
+\int dzdz^\prime\,X_{AC}^{ac}(x,z)\,
\omega^{cC\,dD}(z,z^\prime)\,X_{BD}^{bd}(y,z^\prime)=0\,.
\label{3.15}
\end{equation}

\bigskip\noindent
Using (\ref{3.12}) and (\ref{3.14}), three independent relations from
(\ref{3.15}) are obtained,
\footnote{Detailed reference to the arguments of $X$ and the integral
occurring in (\ref{3.15}) will be subsequently omitted.}

\begin{eqnarray}
A=1,\,B=1&\Longrightarrow&
X_{11}^{ac}X_{12}^{bc}-X_{11}^{bc}X_{12}^{ac}=0\,,
\label{3.16}\\
A=1,\,B=2&\Longrightarrow&
X_{11}^{ac}X_{22}^{bc}-X_{12}^{ac}X_{21}^{bc}
=m^2\,\delta^{ab}\,\delta(x-y)\,,
\label{3.17}\\
A=2,\,B=2&\Longrightarrow&
X_{21}^{ac}X_{22}^{bc}-X_{22}^{ac}X_{21}^{bc}
=-\,gf^{abc}\,(D_i\pi^i)^c\,\delta(x-y)\,.
\label{3.18}
\end{eqnarray}

\bigskip\noindent
As already discussed the number of equations is less than the number
of unknowns. Consequently there is an arbitrariness in the solutions
even after $\omega$ has been fixed. At this point it is important to
carefully choose the solutions for the variables $X$ so that the
subsequent algebra is simplified. From the last equation in the above
set the following choice is practically spelled out,

\begin{eqnarray}
X_{21}^{ac}&=&\delta^{ac}\,\delta(x-z)\,,
\label{3.19}\\
X_{22}^{bc}&=&\frac{1}{2}\,gf^{bcd}\,
(D_i\pi^i)^d\,\delta(x-z)\,.
\label{3.20}
\end{eqnarray}

\bigskip\noindent 
Using (\ref{3.16}), (\ref{3.17}) and the above solutions allows us
to conclude the results for the remaining variables so that the
complete $X$ matrix has the structure,

\begin{equation}
\Bigl(X_{AB}^{ab}(x,y)\Bigr)
=\left(\begin{array}{cc}
0&-\,m^2\,\delta^{ab}\\
\delta^{ab}&\frac{1}{2}\,g\,f^{abc}\,(D_i\pi^i)^c
\end{array}\right)\,\delta(x-y)\,.
\label{3.21}
\end{equation}

\bigskip\noindent
Inserting the expressions for the elements of $X$ in Eq. (\ref{2.9})
immediately leads to the first corrections,

\begin{eqnarray}
T_1^{(1)a}&=&-\,m^2\,\eta^{2a}\,,
\nonumber\\
T_2^{(1)a}&=&\eta^{1a}+\frac{1}{2}\,gf^{abc}\,
\eta^{2b}\,(D_i\pi^i)^c\,,
\nonumber\\
&=&\eta^{1a}+\frac{1}{2}\,g\,\bar\eta^{ac}\,(D_i\pi^i)^c\,,
\label{3.23}
\end{eqnarray}

\bigskip\noindent
where a compact notation is invoked which will prove highly useful,

\begin{equation}
\bar\eta^{ac}=f^{abc}\eta^{2b}\,.
\label{3.23a}
\end{equation}

\bigskip\noindent
The next task is to evaluate the second correction. This can be done
by using Eq. (\ref{2.8b}) which, for the present model, takes the form,

\begin{equation}
\bigl\{T_A^a,T_B^{(1)b}\bigr\}_{(A,\pi)}
+\bigl\{T_A^{(1)a},T_B^b\bigr\}_{(A,\pi)}
+\bigl\{T_A^{(1)a},T_B^{(2)b}\bigr\}_{(\eta)}
+\bigl\{T_A^{(2)a},T_B^{(1)b}\bigr\}_{(\eta)}=0\,.
\label{3.24}
\end{equation}

\bigskip\noindent
Exploiting the expressions for the original constraints (\ref{3.4}),
(\ref{3.8}) and the first corrections (\ref{3.23}) in the above
equation leads to the following conditions for distinct values of $A$
and $B$,

\begin{eqnarray}
&&\underline{A=1,\,B=1}
\nonumber\\
&&\bigl\{\eta^{2a},\,T_1^{(2)b}\bigr\}_{(\eta)}
+\bigl\{T_1^{(2)a},\,\eta^{2b}\bigr\}_{(\eta)}=0\,,
\label{3.25}\\
&&\nonumber\\
&&\underline{A=1,\,B=2}
\nonumber\\
&&m^2\,\bigl\{\eta^{2a},\,T_2^{(2)b}\bigr\}_{(\eta)}
+\bigl\{T_1^{(2)a},\,\eta^{1b}-\frac{1}{2}\,g\,
\bar\eta^{bd}\,(D_i \pi^i)^d\bigr\}_{(\eta)}=0\,,
\label{3.26}\\
&&\nonumber\\
&&\underline{A=2,\,B=2}
\nonumber\\
&&\bigl\{(D_i\pi^i)^a+m^2\,A_0^a,\,\eta^{1b}
+\frac{1}{2}\,g\,\bar\eta^{bd}\,(D_j\pi^j)^d\bigr\}
\nonumber\\
&&\phantom{\bigl\{(D_i\pi^i)^a}
+\bigl\{\eta^{1a}+\frac{1}{2}\,g\,\bar\eta^{ad}\,
(D_i\pi^i)^d,\,(D_j\pi^j)^b+m^2\,A_0^b\bigr\}
\nonumber\\
&&\phantom{\bigl\{(D_i\pi^i)^a}
+\bigl\{\eta^{1a}+\frac{1}{2}\,g\,\bar\eta^{ad}\,
(D_i\pi^i)^d,\,T_2^{(2)b}\bigr\}_{(\eta)}
\nonumber\\
&&\phantom{\bigl\{(D_i\pi^i)^a}
+\bigl\{T_2^{(2)a},\,\eta^{1b}+\frac{1}{2}\,g\,\bar\eta^{bd}\,
(D_j\pi^j)^d\bigr\}_{(\eta)}=0\,.
\label{3.27}
\end{eqnarray}

\bigskip\noindent
The above relation (\ref{3.27}) can be further simplified to yield,

\begin{eqnarray}
&&\bigl\{\eta^{1a}+\frac{1}{2}\,g\,\bar\eta^{ad}\,
(D_i\pi^i)^d,\,T_2^{(2)b}\bigr\}_{(\eta)}
+\bigl\{T_2^{(2)a},\,\eta^{1b}+\frac{1}{2}\,g\,\bar\eta^{bd}\,
(D_j\pi^j)^d\bigr\}_{(\eta)}
\nonumber\\
&=&-\,\frac{1}{2}\,g^2\,f^{abd}\,\bar\eta^{de}\,
(D_i\pi^i)^e\,\delta(x-y)\,.
\label{3.28}
\end{eqnarray}

\bigskip\noindent
It is evident that by choosing the second correction of $T_1^a$ to be
zero achieves considerable simplification. Naturally the same cannot
be done for $T_2^a$ because of (\ref{3.28}). Thus (\ref{3.25}) is
trivially fulfilled while (\ref{3.26}) will also be satisfied
provided $T_2^{(2)a}$ does not depend on $\eta^{1a}$. With this
hypothesis (\ref{3.28}) is solved leading to the final structures,

\begin{eqnarray}
T_2^{(2)a}&=&\frac{g^2}{3!}\bigl(\bar\eta^2\bigr)^{af}\,(D_i\pi^i)^f\,,
\label{3.29}\\
T_1^{(2)a}&=&0\,.
\label{3.30}
\end{eqnarray}

\bigskip\noindent
Note that we have introduced a matrix notation, which will be
frequently used, to denote the product among the $\bar\eta$
variables. It is now useful to explicitly write the involutive
constraints obtained up to the second iterative step. These are given
by,

\begin{eqnarray}
\tilde T_1^a&=&\pi^a_0-m^2\,\eta^{2a}+\cdots
\label{3.31}\\
\tilde T_2^a&=&m^2\,A_0^a+(D_i\pi^i)^a+\eta^{1a}
+\frac{1}{2}\,g\,\bar\eta^{ac}(D_i\pi^i)^c
\nonumber\\
&&\phantom{m^2\,A_0^a+(D_i\pi^i)^a+\eta^{1a}}
+\frac{g^2}{3!}\,\bigl(\bar\eta^2\bigr)^{ae}\,
(D_i\pi^i)^e
+\cdots
\label{3.32}
\end{eqnarray}

\bigskip
It is already seen that while $\tilde T_1^a$ has a simple structure,
the correction terms in $\tilde T_2^a$ generate an exponential-like
series. This is further elaborated by considering the third step of
the method. The equation we have to solve occurs in (\ref{2.8c})
which, in the present instance, reads,

\begin{eqnarray}
&&\bigl\{T_A^a,\,T_B^{(2)b}\bigr\}
+\bigl\{T_A^{(1)a},\,T_B^{(1)b}\bigr\}_{(A,\pi)}
+\bigl\{T_A^{(2)a},\,T_B^b\bigr\}
+\bigl\{T_A^{(1)a},\,T_B^{(3)b}\bigr\}_{(\eta)}
\nonumber\\
&&\phantom{\bigl\{T_A^a,\,T_B^{(2)b}\bigr\}}
+\bigl\{T_A^{(2)a},\,T_B^{(2)b}\bigr\}_{(\eta)}
+\bigl\{T_A^{(3)a},\,T_B^{(1)b}\bigr\}_{(\eta)}=0\,.
\label{3.33}
\end{eqnarray}

\bigskip\noindent
We thus have the following relations for distinct values of $A$ and
$B$, 

\begin{eqnarray}
&&\underline{A=1,\,B=1}
\nonumber\\
&&\bigl\{\eta^{2a},\,T_1^{(3)b}\bigr\}
+\bigl\{T_1^{(3)ab},\,\eta^{2b}\bigr\}=0\,,
\label{3.34}\\
&&\nonumber\\
&&\underline{A=1,\,B=2}
\nonumber\\
&&m^2\,\bigl\{\eta^{2a},\,T_2^{(3)b}\bigr\}
+m^2\,\bigl
\{T_1^{(3)a},\,\eta^{1b}
+\frac{1}{2}\,g\,\bar\eta^{bd}\,(D_i\pi^i)^d
\bigr\}_{(\eta)}=0\,,
\label{3.35}\\
&&\nonumber\\
&&\underline{A=2,\,B=2}
\nonumber\\
&&\bigl\{\eta^{1a}
+\frac{1}{2}\,g\,\bar\eta^{ad}\,(D_i\pi^i)^d,\,
T_2^{(3)b}\bigr\}_{(\eta)}
+\bigl\{T_2^{(3)a},\,
\eta^{1b}+\frac{1}{2}\,g\,\bar\eta^{bd}\,
(D_i\pi^i)^d\bigr\}_{(\eta)}
\nonumber\\
&&\phantom{\bigl\{\eta^{1a}+\frac{1}{2}gf^{acd}}
=g^3\,\Bigl(\frac{1}{6}\,f^{bcd}f^{def}f^{fan}
+\frac{1}{4}\,f^{acd}f^{bef}f^{fdn}
\nonumber\\
&&\phantom{\bigl\{\eta^{1a}+\frac{1}{2}gf^{acd}\,=}
+\frac{1}{6}\,f^{acd}f^{def}f^{bfn}\Bigr)\,
\eta^{2c}\eta^{2e}\,(D_i\pi^i)^n\,\delta(x-y)\,.
\label{3.36}
\end{eqnarray}

\bigskip\noindent
Using our algorithm of taking a vanishing correction for $T_1^a$, it
is simple to observe that these equations will be satisfied by
choosing,

\begin{eqnarray}
T_1^{(3)a}&=&0\,,
\nonumber\\
T_2^{(3)a}&=&\frac{g^3}{4!}\,\bigl(\bar\eta^3\bigr)^{af}\,
(D_i\pi^i)^f\,.
\label{3.36a}
\end{eqnarray}

\bigskip\noindent
This iterative process can be extended to arbitrary orders and the
final expressions for the involutive constraints are given by,

\begin{eqnarray}
\tilde T_1^a&=&\pi_0^a-m^2\,\eta^{2a}\,,
\nonumber\\
\tilde T_2^a&=&m^2\,A_0^a+\eta^{1a}
+\sum_{n=0}^\infty\left[\frac{(g\,\bar\eta)^n}
{(n+1)!}\right]^{ab}(D_i\pi^i)^b\,.
\label{3.36b}
\end{eqnarray}

\bigskip
This completes the first part of the analysis. It is now necessary to
construct the involutive Hamiltonian. We adopt our modified
prescription of first doing this construction for those phase space
variables occurring in the initial canonical Hamiltonian. The
calculation for $\pi_i^a$ is now given here in details. The first
step is to compute the inverses of the matrices (\ref{3.14}) and
(\ref{3.21}), denoted by $(\omega^{ab}_{AB})$ and $(X^{aA,bB})$,
respectively. These are given by,

\begin{equation}
\Bigl(\omega_{AB}^{ab}(x,y)\Bigr)
=\left(\begin{array}{cc}
0&-\,1\\1&0\end{array}\right)\,
\delta^{ab}\delta(x-y)\,,
\label{3.37}
\end{equation}

\bigskip
\begin{equation}
\Bigl(X^{aA,bB}(x,y)\Bigr)
=\left(\begin{array}{cc}
\frac{g}{2m^2}\,f^{abc}\bigl(D_i\pi^i\bigr)^c&\delta^{ab}\\
-\,\frac{1}{m^2}\,\delta^{ab}&0\end{array}\right)\,\delta(x-y)\,.
\label{3.38}
\end{equation}

\bigskip\noindent
Using the elements of these matrices the first correction is obtained
from (\ref{2.16}),

\begin{equation}
\pi_i^{(1)a}=-\,\frac{1}{m^2}\,\eta^{1b}\,G_{1i}^{(0)ba}
+\frac{g}{2m^2}\,f^{bcd}\,\eta^{2b}\,(D_i\pi^i)^c\,
G_{1i}^{(0)da}-\eta^{2b}\,G_{2i}^{(0)ba}\,,
\label{3.39}
\end{equation}

\bigskip\noindent
where the generating functions are defined by,

\begin{eqnarray}
G_{1i}^{(0)ba}&=&\bigl\{T_1^b,\,\pi_i^a\bigr\}=0\,,
\label{3.40}\\
G_{2i}^{(0)ba}&=&\bigl\{T_2^b,\,\pi_i^a\bigr\}
=g\,f^{bad}\,\pi^d_i\,.
\label{3.41}
\end{eqnarray}

\bigskip\noindent
Inserting these in Eq. (\ref{3.39}) the explicit result for the first
correction follows, 

\begin{equation}
\pi_i^{(1)a}=g\,\bar\eta^{ac}\,\pi_i^c\,.
\label{3.42}
\end{equation}

\bigskip\noindent
This will be used to derive the next correction. The result for an
arbitrary iteration is given from the general formula (\ref{2.20}),
which is further simplified by inserting the matrices (\ref{3.37})
and (\ref{3.38})

\begin{eqnarray}
\pi_i^{(n+1)a}&=&-\,\frac{1}{n+1}\,\eta^{Ab}\,
\omega_{AB}^{bc}\,X^{Bc\,Cd}\,G_C^{(n)da}\,,
\nonumber\\
&=&-\,\frac{1}{(n+1)m^2}
\,\eta^{1b}\,G_1^{(n)ba}
+\frac{g}{2\,(n+1)\,m^2}\,\bar\eta^{bc}\,
(D_i\pi^i)^b\,G_1^{(n)ca}
\nonumber\\
&&\phantom{-\,\frac{1}{(n+1)m^2}\,\eta^{1a}\,G_1^{(n)a}}
-\frac{1}{n+1}\,\eta^{2b}\,G_2^{(n)ba}\,.
\label{3.43}
\end{eqnarray}

\bigskip
The structure of the second correction is therefore given by,

\begin{equation}
\pi_i^{(2)a}=-\,\frac{1}{2m^2}\,\eta^{1b}\,G_{1i}^{(1)ba}
+\frac{g}{4m^2}\,\bar\eta^{bc}\,
(D_i\pi^i)^b\,G_{1i}^{(1)ca}
-\frac{1}{2}\,\eta^{2b}\,G_{2i}^{(1)ba}\,,
\label{3.44}
\end{equation}

\bigskip\noindent
where the generating functions, defined in Eq. (\ref{2.21}), are,

\begin{eqnarray}
G_{1i}^{(1)ba}&=&0\,,
\label{3.45a}\\
G_{2i}^{(1)ba}&=&\bigl\{T_2^b,\,\pi_i^{(1)a}\bigr\}
+\bigl\{T_2^{(1)b},\,\pi_i^a\bigr\}
+\bigl\{T_2^{(2)b},\,\pi_i^{(1)a}\bigr\}_{(\eta)}\,,
\nonumber\\
&=&g^2\,f^{bcd}f^{aec}\,\eta^{2e}\pi^d_i
+\frac{g^2}{2}\,f^{bcd}f^{dah}\,\eta^{2c}\pi^h_i\,.
\label{3.45b}
\end{eqnarray}

\bigskip\noindent
Using these results in Eq. (\ref {3.44}), we find the explicit form
of the second correction,

\begin{equation}
\pi_i^{(2)a}=\frac{g^2}{2!}\,(\bar\eta^2)^{ac}\,\pi_i^c\,.
\label{3.46}
\end{equation}

\bigskip\noindent
Similarly, the expression for the third correction looks like,

\begin{equation}
\pi_i^{(3)a}=-\,\frac{1}{3m^2}\,\eta^{1b}\,G_{1i}^{(2)ba}
+\frac{g}{6m^2}\,\bar\eta^{bc}\,(D_j\pi^j)^b\,G_{1i}^{(2)ca}
-\frac{1}{3}\,\eta^{2b}\,G_{2i}^{(2)ba}\,,
\label{3.47}
\end{equation}

\bigskip\noindent
where,

\begin{eqnarray}
G_{1i}^{(2)ba}&=&0\,,
\label{3.48}\\
G_{2i}^{(2)ba}&=&\bigl\{T_2^{(2)b},\,\pi_i^a\bigr\}
+\bigl\{T_2^{(1)b},\,\pi_i^{(1)a}\bigr\}
+\bigl\{T_2^b,\,\pi_i^{(2)a}\bigr\}_{(\eta)}
\nonumber\\
&&+\,\bigl\{T_2^{(2)b},\,\pi_i^{(2)a}\bigr\}_{(\eta)}
+\bigl\{T_2^{(3)b},\,\pi_i^{(1)a}\bigr\}_{(\eta)}\,,
\nonumber\\
&=&\frac{g^3}{6}\,f^{bcd}f^{def}f^{fgh}\,
\eta^{2c}\eta^{2e}\,\bigl\{A_j^g,\,\pi_i^a\bigr\}\,\pi^{hj}
\nonumber\\
&&+\frac{g^3}{2}\,f^{bcd}f^{aef}f^{dgh}\,
\eta^{2c}\eta^{2e}\,\bigl\{A_j^g,\,\pi_i^f\bigr\}\,\pi^{hj}
\nonumber\\
&&+\frac{g^3}{2}\,f^{acd}f^{def}f^{bgh}\,
\eta^{2c}\eta^{2e}\,\bigl\{A_j^g,\,\pi_i^f\bigr\}\,\pi^{hj}\,.
\label{3.49}
\end{eqnarray}

\bigskip\noindent
Using the above results in (\ref{3.47}), we find,

\begin{equation}
\pi_i^{(3)a}=\frac{g^3}{3!}\,(\bar\eta^3)^{ac}\,\pi_i^c\,.
\label{3.50}
\end{equation}

\bigskip\noindent
It is clear that the structure for the corrections to the constraints
has considerably simplified the algebra. There is no correction to
the first generating function while the second acquires a typical
form. This leads to a correction in $\pi_i^a$ which is easily
generalized to arbitrary orders.  The final result for the involutive
expression is thereby given by,

\begin{eqnarray}
\tilde\pi_i^a&=&\pi_i^a+\sum_{n=1}^\infty\pi_i^{(n)a}\,,
\nonumber\\
&=&\sum_{n=0}^\infty\frac{g^n}{n!}\,(\bar\eta^n)^{ac}\,\pi_i^c\,.
\label{3.51}
\end{eqnarray}

\bigskip\noindent
Likewise it is possible to compute the involutive expressions for the
other variables. Since same steps are employed which lead to
identical simplifications, the details are omitted. The final
expression for the involutive $A_i$ is given by,

\begin{eqnarray}
\tilde A_i^a&=&A_i^a+\sum_{n=1}^\infty A_i^{(n)a}\,,
\nonumber\\
&=&A_i^a-g\,\sum_{n=0}^\infty\frac{1}{(n+1)!}\,
(\bar\eta^n)^{ac}\,\bigl(D_i\eta^2\bigr)^c\,.
\label{3.54}
\end{eqnarray}

\bigskip\noindent
The corresponding computation for $A_0$ is slightly tricky since the
normalisation is no longer given by a simple factorial. We quote the
final involutive structure,

\begin{eqnarray}
\tilde A_0^a&=&A_0^a+\sum_{n=1}^\infty A_0^{(n)a}\,,
\nonumber\\
&=&A_0^a+\frac{1}{m^2}\,\sum_{n=1}^\infty
\Bigl[\eta^{1a}\delta_{1n}-g^n\,b_n\,
\bigl(\bar\eta^n\bigr)^{ac}\,\bigl(D_i\pi^i\bigr)^c\Bigr]\,,
\label{3.55}
\end{eqnarray}

\bigskip\noindent
where the coefficients are found from the recursive relation,

\begin{eqnarray}
b_n&=&\frac{1}{n}\,\Bigl[b_{n-1}+\frac{1}{(n+1)!}\Bigr]
\hspace{.5cm}n>1\,,
\nonumber\\
b_1&=&\frac{1}{2}\,.
\label{3.59}
\end{eqnarray}

\bigskip\noindent
The involutive Hamiltonian $\tilde H$ is now obtained directly from
the canonical Hamiltonian that appears in Eq. (\ref{3.6}) by
substituting the initial phase space variables by their corresponding
involutive expressions, i.e.,

\begin{equation}
\tilde H(\tilde A_0,\tilde A_i,\tilde\pi_i)
=H_c(A_0,A_i,\pi_i)_
{\Bigg\vert\begin{array}{c}A_0\rightarrow\tilde A_0\\
A_i\rightarrow\tilde A_i\\
\pi_i\rightarrow\tilde\pi_i\end{array}}\,.
\label{3.60}
\end{equation}

\bigskip\noindent
This completes the BFFT convertion of the second class massive
Yang-Mills theory into a true (first-class) gauge theory. The
constraints (\ref{3.36b}) and Hamiltonian (\ref{3.60}) satisfy a
strongly involutive algebra in the extended phase space.

\vspace{1cm}
\section{Generalized St\"uckelberg formulation}
\renewcommand{\theequation}{4.\arabic{equation}}
\setcounter{equation}{0}

\bigskip
It is well known that a gauge noninvariant Lagrangian can be
converted into a gauge invariant form by introducing auxiliary
scalars. This is the St\"uckelberg \cite{S} formalism by which it is
possible to discuss a gauge invariant formulation of the massive
Maxwell theory, usually called the Proca model.  A non-Abelian
generalisation of the St\"uckelberg mechanism was first proposed by
Kunimasa and Goto \cite{KG}. It was subsequently used to analyse a
gauge invariant Lagrangian formulation of the massive Yang-Mills
theory by a number of authors \cite{SL,SH,CG}. These analyses are
presented either from a geometric or path integral viewpoint.  In
this section we shall first develop in details, using the generalized
St\"uckelberg prescription,  the canonical formalism of the gauge
invariant formulation of the massive Yang-Mills theory in the
coordinate language.  Subsequently it will be shown that the
auxiliary scalars (and their conjugates) introduced in this case are
just the canonical pairs $(\eta^{1a}, \eta^{2a})$ used in the BFFT
approach.

\medskip
The pure (massless) Yang-Mills Lagrangian is invariant under the
gauge transformations,

\begin{equation}
A_\mu^a\longrightarrow\bigl(A_\mu^\theta\bigr)^a
=U^{ab}(\theta)\,A_\mu^b+B_\mu^a(\theta)\,,
\label{4.1}
\end{equation}

\bigskip\noindent 
where $\theta^a(x)$ are the group parameters and,

\begin{equation}
B_\mu^a(\theta)=V^{ab}(\theta)\,\partial_\mu\theta^b\,.
\label{4.2}
\end{equation}

\bigskip\noindent
The transformation matrices $U$ and $V$ are given by,

\begin{eqnarray}
U^{ab}(\theta)&=&\Bigl[\exp\bigl(-\,g\bar\theta\bigr)\Bigr]^{ab}\,,
\label{4.3}\\
V^{ab}(\theta)&=&\sum_{n=0}^\infty
\left[\frac{(-\,g\bar\theta)^n}{(n+1)\,!}\right]^{ab}\,,
\label{4.4}
\end{eqnarray}

\bigskip\noindent
where,

\begin{equation}
\bar\theta^{ab}=f^{acb}\,\theta^c
\label{4.5}
\end{equation}

\bigskip\noindent
uses a notation exploited previously in (\ref{3.23a}). From the above
relations it is simple to deduce the result for infinitesimal
variations,

\begin{equation}
\delta\,A_\mu^\theta=U(\theta)\,\delta\,A_\mu
+D_\mu[A^\theta]\,\left(V(\theta)\,\delta\,\theta\right)
\label{4.6}
\end{equation}

\bigskip\noindent
and $D_\mu$ is the covariant derivative already introduced in
(\ref{3.7}). The inclusion of a mass term in the pure Yang-Mills
Lagrangian, as has been done in Eq. (\ref{3.1}), breaks the
above gauge symmetry. Nevertheless, it is straightforward to
incorporate a non-Abelian gauge invariance into the massive
Yang-Mills Lagrangian by extending the configuration space. This is
the content of the generalized St\"uckelberg mechanism. Using
(\ref{4.2}) and (\ref{4.6}) it can be shown \cite{SH} that by
construction, the following Lagrangian,

\begin{equation}
{\cal L}=-\,\frac{1}{4}\,F_{\mu\nu}F^{\mu\nu}
+\frac{m^2}{2}\,\left(A_\mu-B_\mu(\theta)\right)\,
\left(A^\mu-B^\mu(\theta)\right)
\label{4.7}
\end{equation}

\bigskip\noindent
is invariant under the infinitesimal non-Abelian gauge
transformations, 

\begin{eqnarray}
A_\mu^a&\longrightarrow&A_\mu^a
+D_\mu^{ab}\Bigl(V(\theta)\,\lambda\Bigr)^b\,,
\nonumber\\
\theta^a&\longrightarrow&\theta^a+\lambda^a\,,
\label{4.8}
\end{eqnarray}

\bigskip\noindent
where $\lambda$ is the gauge parameter.

\medskip
Let us next develop the canonical formalism for the gauge invariant
Lagrangian (\ref{4.7}). For subsequent calculations it is useful to
introduce the matrix $W^{ab}(\theta)$ where,

\begin{eqnarray}
W^{ab}(\theta)&=&\sum_{n=1}^\infty
\left[\frac{(-\,g\bar\theta)^n}{(n+1)!}\right]^{ab}\,,
\nonumber\\
&=&V^{ab}(\theta)-\delta^{ab}\,.
\label{4.9}
\end{eqnarray}

\bigskip\noindent
Up to a four divergence the Lagrangian (\ref{4.7}) may be written as,

\begin{eqnarray}
{\cal L}&=&-\,\frac{1}{4}\,F_{\mu\nu}^2
+\frac{m^2}{2}\left(A_\mu^2+B_\mu^2\right)
-m^2\,A_\mu\,\left[\partial^\mu\theta
+W(\theta)\,\partial^\mu\theta\right]\,,
\nonumber\\
&=&-\,\frac{1}{4}\,F_{\mu\nu}^2
+\frac{m^2}{2}\left(A_\mu^2+B_\mu^2\right)
-m^2\,A_\mu\,W(\theta)\partial^\mu\theta
+m^2\,\theta\,\partial_\mu A^\mu\,.
\label{4.10}
\end{eqnarray}

\bigskip\noindent
The canonical momenta are,
\begin{eqnarray}
\pi^a_0&=&\frac{\partial{\cal L}}
{\partial\dot A^{0,a}}=m^2\,\theta^a\,,
\label{4.11}\\
\pi^a_i&=&\frac{\partial{\cal L}}
{\partial\dot A^{i,a}}=-\,F_{0i}^a\,,
\label{4.12}\\
\pi^a_\theta&=&\frac{\partial{\cal L}}
{\partial\dot\theta^a}
=m^2\,V^{ca}V^{cd}\,\dot\theta^d-m^2\,A_0^b\,W^{ba}\,.
\label{4.13}
\end{eqnarray}

\bigskip\noindent
There is one primary constraint,

\begin{equation}
\tilde T_1^a=\pi^a_0-m^2\,\theta^a \approx 0\,.
\label{4.14}
\end{equation}

\bigskip
The above manipulations of isolating the identity component from
$V^{ab}(\theta)$ by defining $W^{ab}(\theta)$ (\ref{4.9}) and then
expressing the Lagrangian up to a four divergence now become clear.
Our motivation was to obtain a constraint that could be identified
with $\tilde T_1^a$ in (\ref{3.36b}). Unless this manipulation was
done the primary constraint obtained from (\ref{4.7}) would just be
$\pi_0^a\approx0$ and the mapping fails.

\medskip
For the computation of the secondary constraint, it is necessary to
obtain the expression for the canonical Hamiltonian. This is given
by,

\begin{eqnarray}
H&=&\int d^3x\,\left(\pi_0\,\dot A^0
+\pi_i\,\dot A^i
+\pi_\theta^a\,\dot\theta^a-{\cal L}\right)
\nonumber\\
&=&H_c+\frac{m^2}{2}\int d^3x\,V^{cd}V^{ca}\,
\dot\theta^d\dot\theta^a+\Delta H\,,
\label{4.15}
\end{eqnarray}

\bigskip\noindent
where $H_c$ is the canonical Hamiltonian that appears in Eq.
(\ref{3.6}) for the usual massive Yang-Mills theory, and,

\begin{equation}
\Delta H=m^2\int d^3x\,\Bigl(\frac{1}{2}\,B^2_i
-\theta^a\,\partial_iA^{i,a}
+A_i^a\,W^{ab}\partial^i\theta^b\Bigr)
\label{4.16}
\end{equation}

\bigskip\noindent
involves terms not depending on either $A_0$ or $\dot\theta$ so that
it does not influence the computation of the secondary constraint. In
order to simplify the velocity dependent term in (\ref{4.15}) the
first step is to invert (\ref{4.13}) so that the velocity
$\dot\theta^a$ is expressed in terms of the momenta $\pi^a_\theta$.
This is done in Appendix A. Next, using the results of this Appendix,
the desired simplification is done in Appendix B.  Using (\ref{B.4})
in (\ref{4.15}) the final form of the Hamiltonian is obtained,

\begin{equation}
H=H_c+\Delta H
+\frac{m^2}{2}\int d^3x\,\Gamma^a\Gamma^g\sum_{n=0}^\infty(-1)^n\,
\left(\Lambda^n\right)^{ga}\,.
\label{4.17}
\end{equation}

\bigskip\noindent
Time conserving the primary constraint yields the secondary
constraint $\tilde T_2^a$,

\begin{equation}
\left\{\tilde T_1^a,\,H\right\}
=\left\{\pi_0^a-m^2\,\theta^a,\,H\right\}
=\tilde T_2^a\approx0\,,
\label{4.18}
\end{equation}

\bigskip\noindent
where,

\begin{equation}
\tilde T_2^a=T_2^{a}+\Delta T^a
\label{4.19}
\end{equation}

\bigskip\noindent
involves the sum of the secondary constraint (\ref{3.8}) occurring in
the usual massive Yang-Mills theory and an extra piece,

\begin{equation}
\Delta T^a=-\,\pi_\theta^a-\pi_\theta^d\,\chi^{da}
+m^2\,A_0^d\,\Omega^{da}\,.
\label{4.20}
\end{equation}

\bigskip\noindent
The structures $\chi$ and $\Omega$ have been defined as well as
simplified in Appendix C (\ref{C.15} and \ref{C.26}).

\medskip
We now make certain observations. An inspection shows that the
constraint (\ref{4.14}) gets mapped on to the first constraint in
(\ref{3.36b}) by a simple identification of fields. However the same
thing cannot be done for the other constraint. Nevertheless, as we
shall now show by a series of manipulations, a correspondence between
these constraints can also be achieved.  Using (\ref{C.26}) in
(\ref{4.20}) we obtain,

\begin{equation}
\Delta T^a=\Theta^d\,\Omega^{da}-\pi_\theta^a\,,
\label{4.21}
\end{equation}

\bigskip\noindent
where

\begin{equation}
\Theta^d=-\pi_\theta^d
+m^2\,A_0^d\,.
\label{4.22}
\end{equation}

\bigskip\noindent
On the constraint surface $\tilde T_2^a=0$ we find the
solution,\footnote{It is crucial to note that the equality between
$\chi$ and $\Omega$ is essential to derive this result, which plays a
central role in the forthcoming analysis.}

\begin{eqnarray}
\Theta^a&=&-\,\left(D_i\pi^i\right)^d\sum_{n=0}^\infty(-1)^n\,
\left[\Omega^n\right]^{da}\,,
\nonumber\\
&=&-\,\left(D_i\pi^i\right)^d\,
\left(\delta^{da}+\Omega^{da}\right)^{-1}\,.
\label{4.23}
\end{eqnarray}

\bigskip\noindent
Defining the inverse by $I$ so that,

\begin{equation}
\left(\delta^{ab}+\Omega^{ab}\right)\,I^{bc}=\delta^{ac}\,.
\label{4.24}
\end{equation}

\bigskip\noindent
Expressing this as a series,

\begin{equation}
I^{ab}=\delta^{ab}+\sum_{n=1}^\infty\beta_n\,
\left[(g\bar\theta)^n\right]^{ab}
\label{4.24b}
\end{equation}

\bigskip\noindent
and inserting it in (\ref{4.24}) leads to the following condition
for $\beta_n$;

\begin{equation}
\sum_{n=1}^\infty\beta_n\,
\left[(g\bar\theta)^n\right]^{ab}
+\Omega^{ab}+\Omega^{ac}\sum_{n=1}^\infty\beta_n\,
\left[(g\bar\theta)^n\right]^{cb}=0\,.
\label{4.25}
\end{equation}

\bigskip\noindent
From the above relations, therefore, the extra piece $\Delta T^a$
simplifies to

\begin{eqnarray}
\Delta T^a&=&-\,\left(D_i\pi^i\right)^e\,\Omega^{eb}\,I^{ba}
-\pi^a_\theta\,,
\nonumber\\
&=&\left(D_i\pi^i\right)^e\sum_{n=0}^\infty\beta_n\,
\left[(g\bar\theta)^n\right]^{ea}-\pi^a_\theta\,.
\label{4.26}
\end{eqnarray}

\bigskip\noindent
The final task is to compute $\beta_n$ by solving Eq. (\ref{4.25}).
This is done in Appendix D. Using Eq. (\ref{D.4}) the final
expression for the constraints is given,

\begin{eqnarray}
\tilde T_1^a&=&T_1^a-m^2\theta^a\,,
\nonumber\\
&=&\pi_0^a-m^2\theta^a\,, 
\nonumber\\
\tilde T_2^a&=&T_2^a-\pi_\theta^a
+\sum_{n=1}^\infty a_n\,
\left[(g\bar\theta)^n\right]^{ba}\,\left(D_i\pi^i\right)^b\,,
\nonumber\\
&=&m^2\,A_0^a-\pi_\theta^a+\sum_{n=0}^\infty a_n\,
\left[(g\bar\theta)^n\right]^{ba}\,\left(D_i\pi^i\right)^b\,,
\nonumber\\
&=&m^2\,A_0^a-\pi_\theta^a+\sum_{n=0}^\infty\,
\left[\frac{(g\bar\theta)^n}{(n+1)!}\right]^{ab}\,
\left(D_i\pi^i\right)^b\,,
\label{4.27}
\end{eqnarray}

\bigskip\noindent
where (\ref{A.5a}) has been used to obtain the last line.  If we make
the following identifications with the fields introduced in the BFFT
approach,

\begin{equation}
\theta^a\longleftrightarrow\eta^{2a},
\hspace{.5cm}
\pi^a_\theta\longleftrightarrow-\,\eta^{1a}
\label{4.28}
\end{equation}

\bigskip\noindent
the canonical algebra is preserved,

\begin{equation}
\left\{\theta^a(x),\,\pi_\theta^b(y)\right\}
=-\,\left\{\eta^{2a}(x),\,\eta^{1b}(y)\right\}
=\delta^{ab}\,\delta(x-y)\,.
\label{4.29}
\end{equation}

\bigskip\noindent
Now the above results for $\tilde T_1^a$ and $\tilde T_2^a$ are
exactly identifiable with the corresponding expressions given in
(\ref{3.36b}). This shows that the auxiliary scalars in the
generalized St\"uckelberg formalism are exactly mapped on to the BFFT
fields.

\bigskip
Before closing this section it may be worthwhile to point out that
the constraint in the form (\ref{4.27}) (and not as it occurs in
(\ref{4.20})) reveals that it is the generator of gauge
transformations because,

\begin{eqnarray}
\int d^3x\,\bigl\{\tilde T_2^a(x)\,\lambda^a(x),\,\theta^b(y)\bigr\}
&=&\lambda^b(y)\,,
\nonumber\\
\int d^3x\,\bigl\{\tilde T_2^a(x)\,\lambda^a(x),\,A_i^b(y)\bigr\}
&=&D_i^{bd}\,\left(V^{da}\,\lambda^a(y)\right)\,,
\label{4.30}
\end{eqnarray}

\bigskip\noindent
which correctly reproduces (\ref{4.8}).

\vspace{1cm}
\section{Conclusion}

\bigskip
We have discussed a systematic method, within the Batalin, Fradkin,
Fradkina and Tyutin (BFFT) \cite{B,BT} approach, of converting the
massive Yang- Mills theory to a gauge invariant theory by extending
the phase space.  Exploiting an intelligent choice for the symplectic
matrix $\omega^{ab}$ and the generating matrix $X^{ab}$, the infinite
number of iterative corrections necessary for obtaining the strongly
involutive constraints were considerably simplified. These
corrections were explicitly computed and expressed in a closed
(exponential-like) form. In obtaining the strongly involutive
Hamiltonian, on the other hand, the conventional BFFT approach was
modified. First, the strongly involutive forms for the initial phase
space variables were deduced. Once again infinite sets of iterative
corrections were necessary which were computed and put in
exponential-like series. Then the canonical Hamiltonian of the
massive Yang-Mills theory was rewritten, replacing the original phase
space variables by their corresponding involutive expressions. This
directly gave the cherished form of the involutive Hamiltonian.
Indeed any dynamical variable in the original theory can be converted
into its involutive form by this technique.

\medskip
In the latter half of the paper the generalized St\"uckelberg
formalism \cite{KG} of changing the gauge noninvariant massive
Yang-Mills Lagrangian into a gauge invariant form by introducing
auxiliary scalars was reconsidered.  The complete canonical formalism
of the gauge invariant Lagrangian was developed in the coordinate
basis. The explicit structures of the involutive constraints and
Hamiltonian were determined. Subsequently it was shown by a series of
algebraic simplifications that the auxiliary St\"uckelberg scalars
and their conjugates were exactly identified with the additional
canonical pairs of fields, defined in the extended phase space,
invoked earlier in the context of the BFFT analysis. By this
identification a mapping between the Lagrangian and Hamiltonian
embedding prescriptions of transforming gauge noninvariant into gauge
invariant systems based on the generalized St\"uckelberg and the BFFT
approaches, respectively, was established. It should be stressed that
this mapping is independent of either the space-time dimensionality
or the specific non-Abelian gauge group employed in the analysis. In
this sense, therefore, the present paper extended and generalized
similar correspondences reported earlier \cite{X,BBG} for abelian
groups. In this instance the abelian result follows trivially by
setting the structure constants to zero.

\medskip
It is clear that just as the generalisation of the usual
St\"uckelberg formalism \cite{S} from Abelian to non-Abelian
theories \cite{KG} is nontrivial, the same is true in the BFFT
formalism.  Indeed a distinctive feature of transforming non-Abelian
second class into first class systems, in contrast to abelian
theories, was that all orders of iteration in the BFFT approach were
mandatory. But our approach provided an algorithm of systematically
computing these corrections, and expressing them in closed forms. An
interesting application of the method developed here would be to
study the bosonisation and duality among non-Abelian theories in
higher dimensions where the conventional master Lagrangian approach
\cite{K1} had failed whereas, that based on the generalised
St\"uckelberg formalism was effective \cite{BBG2}.

\medskip
As final remarks we mention that the Hamiltonian formalism for the
massive Yang-Mills theory, regarded as a second class system, was
originally developed in \cite{Senj}. The corresponding first class
(canonical) interpretation in terms of the St\"uckelberg approach was
earlier presented in \cite{CG} which, contrary to the present work,
starts from an embedded Lagrangian and does not use the elaborate
BFFT method.  Incidentally, the idea of converting second class
systems their first class forms by extending the phase space, which
is the crux of this method, was initially suggested in \cite{FS}.

\vspace{1cm}
\noindent
{\bf Acknowledgment:} This work is supported in part by Conselho
Nacional de Desenvolvimento Cient\'{\i}fico e Tecnol\'ogico - CNPq,
Financiadora de Estudos e Projetos - FINEP and Funda\c{c}\~ao
Universit\'aria Jos\'e Bonif\'acio - FUJB (Brazilian Research
Agencies). One of the authors (RB) would like to thank the members of
the Department of Theoretical Physics, UFRJ, for their kind
hospitality.

\vspace{1cm}
\appendix
\renewcommand{\theequation}{A.\arabic{equation}}
\setcounter{equation}{0}
\section*{Appendix A}

\bigskip
This Appendix is devoted to invert (\ref{4.13}) so that the
velocities are expressed in terms of the momenta. The product of the
$V$ matrices 
\footnote{In a geometric language this product has the meaning of a
two sided invariant metric on the group of rotations \cite{SL}.}
is simplified by using (\ref{4.4}),

\begin{eqnarray}
V^{ca}(\theta)\,V^{cb}(\theta)
&=&\sum_{n=0}^\infty\left[\frac{(g\bar\theta)^{2n}}
{(n+1)(2n+1)!}\right]^{ab}\,,
\nonumber\\
&=&\bar V^{ab}(\theta)
=\bar V^{ba}(\theta)\,,
\label{A.1}
\end{eqnarray}

\bigskip\noindent
where the symmetry follows from the definition of $\bar\theta$ given
in Eq. (\ref{4.5}) since,

\begin{equation}
\bigl[\bar\theta^n\bigr]^{ab}=(-1)^n\,
\bigl[\bar\theta^n\bigr]^{ba}\,.
\label{A.5a}
\end{equation}

\bigskip\noindent

It is once again useful to separate the $n=0$ contribution,

\begin{eqnarray}
\bar V^{ab}(\theta)&=&\delta^{ab}+\Lambda^{ab}(\theta)\,,
\label{A.2}\\
\Lambda^{ab}(\theta)
&=&\sum_{n=1}^\infty\left[\frac{(g\bar\theta)^{2n}}
{(n+1)(2n+1)!}\right]^{ab}
=\Lambda^{ba}(\theta)\,.
\label{A.3}
\end{eqnarray}

\bigskip\noindent
Defining

\begin{equation}
\Gamma^a=\frac{\pi_\theta^a}{m^2}+A_0^b\,W^{ba}
\label{A.4}
\end{equation}

\bigskip\noindent
and using the above relations enables Eq. (\ref{4.13}) to be
expressed as,

\begin{equation}
\Gamma^a=\dot\theta^a+\dot\theta^b\,\Lambda^{ba}\,.
\label{A.5}
\end{equation}

\bigskip
The solution for $\dot\theta^a$ is now given by,

\begin{equation}
\dot\theta^a=\Gamma^b\sum_{n=0}^\infty(-\,1)^n\,
\left(\Lambda^n\right)^{ba}\,,
\label{A.6}
\end{equation}

\bigskip\noindent
which may be verified by directly substituting in the RHS of
(\ref{A.5}),

\begin{equation}
\Gamma^b\sum_{n=0}^\infty(-\,1)^n\left(\Lambda^n\right)^{ba}
+\Gamma^b\sum_{n=0}^\infty(-\,1)^n\left(\Lambda^{n+1}\right)^{ba}\,.
\label{A.7}
\end{equation}

\bigskip\noindent
Changing the sum in the second term from $n$ to $m=n+1$ leads to

\begin{equation}
\Gamma^b\sum_{n=0}^\infty(-\,1)^n\left(\Lambda^n\right)^{ba}
-\Gamma^b\sum_{m=1}^\infty(-\,1)^m\left(\Lambda^m\right)^{ba}
=\Gamma^a\,,
\label{A.8}
\end{equation}

\bigskip\noindent
which reproduces the LHS of (\ref{A.5}).

\vspace{1cm}
\appendix
\renewcommand{\theequation}{B.\arabic{equation}}
\setcounter{equation}{0}
\section*{Appendix B}

\bigskip
The velocity dependent factor in the Hamiltonian (\ref{4.15}) will be
expressed here in terms of the phase space variables. This factor is
written as

\begin{eqnarray}
V^{ca}V^{cb}\,\dot\theta^a\dot\theta^b
&=&\left(\delta^{ab}+\Lambda^{ab}\right)\,
\dot\theta^a\dot\theta^b\,,
\nonumber\\
&=&\left(\delta^{ab}+\Lambda^{ab}\right)\,
\Gamma^c\sum_{n=0}^\infty(-1)^n\,\left(\Lambda^n\right)^{ca}\,
\Gamma^d\sum_{m=0}^\infty(-1)^m\,\left(\Lambda^m\right)^{db},
\label{B.1}
\end{eqnarray}

\bigskip\noindent
where we have used (\ref{A.1}), (\ref{A.2}) and (\ref{A.6}). The RHS
simplifies to,

\begin{eqnarray}
&&\Gamma^c\sum_{n=0}^\infty(-1)^n\,\left(\Lambda^n\right)^{cb}\,
\Gamma^d\sum_{m=0}^\infty(-1)^m\,\left(\Lambda^m\right)^{db}
\nonumber\\
&&+\Gamma^c\sum_{n=0}^\infty(-1)^n\,\left(\Lambda^{n+1}\right)^{cb}\,
\Gamma^d\sum_{m=0}^\infty(-1)^m\,\left(\Lambda^m\right)^{db}\,.
\label{B.2}
\end{eqnarray}

\bigskip\noindent
Consider the second term. Replace the first sum from $n$ to
$m^\prime=n+1$,

\begin{equation}
-\,\Gamma^c\sum_{m^\prime=1}^\infty(-1)^{m^\prime}\,
\left(\Lambda^{m^\prime}\right)^{cb}\,
\Gamma^d\sum_{m=0}^\infty(-1)^m\,\left(\Lambda^m\right)^{db}\,.
\label{B.3}
\end{equation}

\bigskip\noindent
Adding this with the first term in Eq. (\ref{B.2}) yields the final
result 

\begin{equation}
V^{ca}V^{cb}\,\dot\theta^a\dot\theta^b
=\Gamma^a\Gamma^b\sum_{m=0}^\infty(-1)^m\,
\left(\Lambda^m\right)^{ab}\,.
\label{B.4}
\end{equation}

\vspace{1cm}
\appendix
\renewcommand{\theequation}{C.\arabic{equation}}
\setcounter{equation}{0}
\section*{Appendix C}

\bigskip
The matrices $\chi$ and $\Omega$ appearing in the constraint
(\ref{4.19}) are considered here in details. We first treat $\chi$
whereas $\Omega$ is taken up from (\ref{C.17}) onwards. The matrix
elements of $\chi$ are given by,

\begin{equation}
\chi^{ab}=\sum_{n=1}^\infty\left(-1\right)^n\,
\left(\Lambda^n\right)^{ab}
+W^{ac}\sum_{n=0}^\infty\left(-1\right)^n\,
\left(\Lambda^n\right)^{cb}\,,
\label{C.1}
\end{equation}

\bigskip\noindent
where $W$ and $\Lambda$ have been introduced in (\ref{4.9}) and
(\ref{A.3}), respectively.  Consider the expression,

\begin{equation}
\sum_{n=0}^\infty\left(-1\right)^n\,
\left(\Lambda^n\right)^{ab}
=\left(\delta^{ab}+\Lambda^{ab}\right)^{-1}\,.
\label{C.2}
\end{equation}

\bigskip\noindent
Writing this as a series in $\theta$,

\begin{equation}
\sum_{n=0}^\infty\left(-1\right)^n\,
\left(\Lambda^n\right)^{ab}
=\sum_{n=0}^\infty C_n\,
\left[\left(g\bar\theta\right)^n\right]^{ab}\,,
\label{C.3}
\end{equation}

\bigskip\noindent
it is possible to determine $C_n$ using (\ref{C.2}),

\begin{equation}
\left(\delta^{ac}+\Lambda^{ac}\right)\,
\sum_{n=0}^\infty C_n\,
\left[\left(g\bar\theta\right)^n\right]^{cb}
=\delta^{ab}\,.
\label{C.4}
\end{equation}

\bigskip\noindent
It is easy to verify that only even powers of $(g\bar\theta)$
contribute.  The final result is

\begin{equation}
\sum_{n=0}^\infty\left(-1\right)^n\,
\left(\Lambda^n\right)^{ab}
=\sum_{n=0}^\infty c_n\,
\left[\left(g\bar\theta\right)^{2n}\right]^{ab}\,,
\label{C.5}
\end{equation}

\bigskip\noindent
where

\begin{eqnarray}
c_0&=&1\,,
\nonumber\\
c_n&=&-\,\sum_{m=1}^nA_m\,c_{n-m}
\hspace{.5cm}(n>0)\,,
\nonumber\\
A_m&=&\frac{1}{(m+1)(2m+1)!}\,.
\label{C.6}
\end{eqnarray}

\bigskip\noindent
We now simplify the second term in (\ref{C.1});

\begin{equation}
W^{ac}\sum_{n=0}^\infty(-1)^n\,
\left(\Lambda^n\right)^{cb}
=\sum_{p=1}^\infty a_p\,
\left[(g\bar\theta)^p\right]^{ac}
\sum_{n=0}^\infty c_n\,
\left[(g\bar\theta)^{2n}\right]^{cb}\,,
\label{C.7}
\end{equation}

\bigskip\noindent
with,

\begin{equation}
a_p=\frac{(-1)^p}{(p+1)!}\,.
\label{C.8}
\end{equation}

\bigskip\noindent
Let us express the product occurring in the RHS of (\ref{C.7}) as a
power series where the odd and even terms are explicitly separated,

\begin{equation}
-\,\frac{g\bar\theta^{ab}}{2!}
+\sum_{n=1}^\infty d_n\,
\left[(g\bar\theta)^{2n}\right]^{ab}
+\sum_{n=1}^\infty f_n\,
\left[(g\bar\theta)^{2n+1}\right]^{ab}\,.
\label{C.9}
\end{equation}

\bigskip\noindent
A straightforward comparison of terms fixes $d_n$ and $f_n$ to be,

\begin{eqnarray}
d_n&=&\sum_{m=0}^{n-1}a_{2(n-m)}\,c_m\,,
\nonumber\\
f_n&=&\sum_{m=0}^na_{2(n-m)+1}\,c_m\,.
\label{C.10}
\end{eqnarray}

\bigskip\noindent
Writing the complete expansion for $f_n$ yields,

\begin{eqnarray}
f_n&=&a_{2n+1}\,c_0+a_{2n-1}\,c_1
+\cdots+a_1\,c_n\,,
\nonumber\\
&=&a_{2n+1}\,c_0+a_{2n-1}\,c_1
+\cdots-a_1\,\left(A_1\,c_{n-1}
+\cdots+A_n\,c_0\right)\,,
\label{C.11}
\end{eqnarray}

\bigskip\noindent
where we have inserted the expansion of $c_n$ in the last term. Now
using the identity

\begin{equation}
a_1\,A_n=a_{2n+1}\,,
\label{C.12}
\end{equation}

\bigskip\noindent
that follows from the respective definitions (\ref{C.6}) and
(\ref{C.8}), it is seen that the last factor in the parenthesis
cancels the first term in $f_n$.  Likewise, there will be a pair wise
cancellations among all terms so that,

\begin{equation}
f_n=0\,.
\label{C.13}
\end{equation}

\bigskip\noindent
Collecting all pieces together, we obtain,

\begin{equation}
\chi^{ab}=\sum_{n=1}^\infty(-1)^n\,
\left(\Lambda^n\right)^{ab}
-\frac{g\bar\theta^{ba}}{2!}
+\sum_{n=1}^\infty d_n\,
\left[(g\bar\theta)^{2n}\right]^{ba}\,.
\label{C.14}
\end{equation}

\bigskip\noindent
Using the result (\ref{C.5}) and the symmetry properties of
$\bar\theta^{ba}$, $\chi^{ab}$ is further simplified,

\begin{equation}
\chi^{ab}=\frac{g}{2!}\,\bar\theta^{ab}
+\sum_{n=1}^\infty\omega_n\,
\left[(g\bar\theta)^{2n}\right]^{ab}\,,
\label{C.15}
\end{equation}

\bigskip\noindent
where,

\begin{eqnarray}
\omega_n&=&c_n+d_n\,,
\nonumber\\
&=& \sum_{m=0}^na_{2(n-m)}\,c_m\,,
\label{C.16}
\end{eqnarray}

\bigskip\noindent
which follows from the explicit expression for $d_n$ given in
(\ref{C.10}) and $a_0=1$. 

\medskip
We now consider the matrix $\Omega$ which is defined by,

\begin{eqnarray}
\Omega^{ab}&=&-\,W^{ac}\sum_{n=0}^\infty(-1)^n\,
\left(\Lambda^n\right)^{cb}
-W^{ac}W^{bd}\sum_{n=0}^\infty(-1)^n\,
\left(\Lambda^n\right)^{dc}\,,
\nonumber\\
&=&-\,W^{ac}\,\left\{\delta^{bc}
-\frac{g\bar\theta^{bc}}{2}
+\sum_{n=1}^\infty\omega_n\,
\bigl[(g\bar\theta)^{2n}\bigr]^{bc}\right\}\,,
\label{C.17}
\end{eqnarray}

\bigskip\noindent
where the second line follows from the previous analysis in this
Appendix.  Putting the explicit form for $W$ leads to,

\begin{eqnarray}
\Omega^{ab}&=&-\,\Bigl\{\sum_{p=1}^\infty a_p\,
\bigl[(g\bar\theta)^p\bigr]^{ab}
+\frac{1}{2}\sum_{p=1}^\infty a_p\,
\bigl[(g\bar\theta)^{p+1}\bigr]^{ab}
\nonumber\\
&&\phantom{\sum_{p=1}^\infty a_p\,\bigl[g\bigr]}
+\sum_{p=1}^\infty a_p\,
\bigl[(g\bar\theta)^p\bigr]^{ac}
\sum_{n=1}^\infty\omega_n\,
\bigl[(g\bar\theta)^{2n}\bigr]^{cb}\Bigr\}\,.
\label{C.18}
\end{eqnarray}

\bigskip\noindent
The first two terms are combined by a simple change in the summation
$p\longrightarrow p+1$. It leads to,

\begin{eqnarray}
\Omega^{ab}&=&-\,\sum_{p=1}^\infty a_p\,
\bigl[(g\bar\theta)^p\bigr]^{ac}
\sum_{n=1}^\infty\omega_n\,
\bigl[(g\bar\theta)^{2n}\bigr]^{cb}
\nonumber\\
&&-\,\sum_{q=1}^\infty\frac{q}{2(q+2)}\,a_q\,
\bigl[(g\bar\theta)^{q+1}\bigr]^{ab}
+\frac{g\bar\theta^{ab}}{2}\,.
\label{C.19}
\end{eqnarray}

\bigskip\noindent
We now prove that terms involving odd powers of $g$ in the two
summations conspire to vanish. An arbitrary odd powered term
$(g\bar\theta)^{2n+1}\hspace{.3cm}(n=1,\,2,\,\dots)$ has the
coefficient (apart from a minus sign),

\begin{equation}
p_{2n+1}=a_{2n-1}\,\omega_1+a_{2n-3}\,\omega_2
+\cdots+a_3\,\omega_{n-1}+a_1\,\omega_n
+\frac{n}{2n+2}\,a_{2n}\,.
\label{C.20}
\end{equation}

\bigskip\noindent
Inserting the expansion for $\omega_n$, given in (\ref{C.16}), this
yields,

\begin{eqnarray}
p_{2n+1}&=&a_{2n-1}\,\left(a_2c_0+a_0c_1\right)
+a_{2n-3}\,\left(a_4c_0+a_2c_1+a_0c_2\right)
\nonumber\\
&&+\cdots+a_1\,\left(a_{2n}c_0+a_{2n-2}c_1+\cdots+a_0c_n\right)
+\frac{n}{2n+2}\,a_{2n}\,.
\label{C.21}
\end{eqnarray}

\bigskip\noindent
The term $a_1a_0c_n$ in the last parenthesis is further elaborated by
decomposing $c_n$,

\begin{equation}
a_1a_0c_n=-\,a_1a_0\,\left(A_1c_{n-1}+A_2c_{n-2}
+\cdots A_{n-1}c_1+A_nc_0\right)\,.
\label{C.22}
\end{equation}

\bigskip\noindent
Using the identity (\ref{C.12}), it is found that the penultimate
term in (\ref{C.22}) cancels the second factor in the first
parenthesis of (\ref{C.21}). Indeed, just as discussed in Appendix B,
there will be pair wise cancellation except for the last factor in
(\ref{C.22}) and the first one in the final parenthesis of
(\ref{C.21}). Combining these remaining pieces,

\begin{equation}
p_{2n+1}=a_1\,a_{2n}\,c_0-a_1\,a_0\,A_n\,c_0
+\frac{n}{2n+2}\,a_{2n}=0\,,
\label{C.23}
\end{equation}

\bigskip\noindent
which follows from the explicit expressions. Hence, as announced
before, terms with odd powers of $g$ vanish. Next, looking at even
powered terms, the coefficient of
$O\,(g^{2n})\hspace{.3cm}(n=1,\,2,\,\dots)$ is given by,

\begin{equation}
p_{2n}=-\left\{a_{2n-2}\,\omega_1+a_{2n-4}\,\omega_2
+\cdots+a_4\,\omega_{n-2}+a_2\,\omega_{n-1}\right\}
-\frac{2n-1}{2(2n+1)}\,a_{2n-1}\,.
\label{C.24}
\end{equation}

\bigskip\noindent
By following analogous steps it may be shown
\footnote{We shall subsequently discuss in Appendix D (\ref{D.8} to
\ref{D.11}) how this result also follows from a different viewpoint.} 
that this simplifies to

\begin{equation}
p_{2n}=\sum_{m=0}^na_{2(n-m)}\,c_m=\omega_n\,.
\label{C.25}
\end{equation}

\bigskip\noindent
Hence we find the remarkable result,
\begin{eqnarray}
\Omega^{ab}&=&\frac{g\bar\theta^{ab}}{2}
+\sum_{n=1}^\infty\omega_n\,
\bigl[(g\bar\theta)^{2n}\bigr]^{ab}\,,
\nonumber\\
&=&\chi^{ab}\,.
\label{C.26}
\end{eqnarray}

\vspace{1cm}
\appendix
\renewcommand{\theequation}{D.\arabic{equation}}
\setcounter{equation}{0}
\section*{Appendix D}

\bigskip
The explicit solution for $\beta_n$ will be obtained by first
expressing (\ref{4.25}) in full form, 

\begin{eqnarray}
&&\sum_{n=1}^\infty\beta_n\,\left[(g\bar\theta)^n\right]^{ab}
+\frac{g\bar\theta^{ab}}{2!}
+\sum_{n=1}^\infty\omega_n\,\left[(g\bar\theta)^{2n}\right]^{ab}
\nonumber\\
&&\phantom{\sum_{n=1}^\infty\beta_n\,\left[(g\bar\theta)^n\right]^{ab}}
+\left(\frac{g\bar\theta^{ac}}{2}
+\sum_{n=1}^\infty\omega_n\,\left[(g\bar\theta)^{2n}\right]^{ac}
\right)\sum_{n=1}^\infty\beta_n\,\left[(g\bar\theta)^n\right]^{cb}
=0.
\label{D.1}
\end{eqnarray}

\bigskip\noindent
By equating coefficients of the first few terms, we find,

\begin{eqnarray}
\beta_1+\frac{1}{2!}&=&0\,,
\nonumber\\
\beta_2+\omega_1+\frac{\beta_1}{2!}&=&0\,,
\nonumber\\
\beta_3+\frac{\beta_2}{2!}+\omega_1\beta_1&=&0\,,
\nonumber\\
\beta_4+\omega_2+\frac{\beta_3}{2!}+\omega_1\beta_2&=&0\,,
\nonumber\\
&\vdots&
\label{D.2}
\end{eqnarray}

\bigskip\noindent
The first equation gives $\beta_1$ which is used to obtain $\beta_2$
from the next equation. This iterative process can be continued to
obtain all the $\beta$'s. Indeed an explicit computation, using the
value of $\omega_n$ from (\ref{C.16}), shows

\begin{eqnarray}
\beta_1&=&-\,\frac{1}{2!}\,,
\nonumber\\
\beta_2&=&\frac{1}{3!}\,,
\nonumber\\
\beta_3&=&-\,\frac{1}{4!}\,,
\nonumber\\
\beta_4&=&\frac{1}{5!}\,,
\nonumber\\
&\vdots&
\label{D.3}
\end{eqnarray}

\bigskip\noindent
which suggests a general solution for $\beta_n$;

\begin{equation}
\beta_n=\frac{(-1)^n}{(n+1)!}=a_n\,.
\label{D.4}
\end{equation}

\bigskip\noindent
It is now simple to prove this result explicitly. Take the
coefficient of the $O(g^{2n+1})$ term ($n=1,\,2,\,\cdots$),

\begin{equation}
\beta_{2n+1}+\frac{\beta_{2n}}{2!}
+\sum_{m=1}^n\omega_m\,\beta_{2(n-m)+1}=0\,.
\label{D.5}
\end{equation}

\bigskip\noindent
Setting the solution (\ref{D.4}) for $\beta_n$, the LHS of the above
equation yields,

\begin{eqnarray}
&&a_{2n+1}+\frac{a_{2n}}{2!}
+\sum_{m=1}^n\omega_m\,a_{2(n-m)+1}
\nonumber\\
&=&\left(a_{2n+1}+\frac{a_{2n}}{2}\right)
+\omega_1\,a_{2n-1}+\omega_2\,a_{2n-3}
+\cdots+\omega_n\,a_1\,.
\label{D.6}
\end{eqnarray}

\bigskip\noindent
The series separated from the parenthesis has already been evaluated
in Appendix C (see \ref{C.20} and \ref{C.23}). Inserting this result
in (\ref{D.6}) we obtain,

\begin{equation}
a_{2n+1}+\frac{a_{2n}}{2}-\frac{n}{2n+2}\,a_{2n}=0\,,
\label{D.7}
\end{equation}

\bigskip\noindent
obtained by using (\ref{C.8}). Thus the consistency of the solution
for $\beta_n$ is verified.

\medskip
It is instructive to observe the coefficients of the $O(g^{2n})$ term
$(n=2,\,3,\,\dots)$ which is given by,

\begin{equation}
\beta_{2n}+\omega_n+\frac{\beta_{2n-1}}{2!}
+\sum_{m=1}^{n-1}\omega_m\,\beta_{2(n-m)}=0\,.
\label{D.8}
\end{equation}

\bigskip\noindent
Inserting the solution for $\beta_n$, it follows,

\begin{equation}
-\sum_{m=1}^{n-1}\omega_m\,a_{2(n-m)}-a_{2n}
-\frac{a_{2n-1}}{2!}=\omega_n\,.
\label{D.9}
\end{equation}

\bigskip\noindent
It may be easily checked that,

\begin{equation}
a_{2n}+\frac{a_{2n-1}}{2!}
=\frac{2n-1}{2(2n+1)}\,a_{2n-1}
\label{D.10}
\end{equation}

\bigskip\noindent
and we obtain,

\begin{equation}
-\sum_{m=1}^{n-1}\omega_m\,a_{2(n-m)}
-\frac{2n-1}{2(2n+1)}\,a_{2n-1}=\omega_n\,,
\label{D.11}
\end{equation}

\bigskip\noindent
which just reproduces the result (\ref{C.24}) and (\ref{C.25})
\footnote{Refer to footnote 6}.

\end{document}